\documentclass[fleqn,usenatbib]{mnras}

\usepackage{newtxtext,newtxmath}
\usepackage[T1]{fontenc}
\usepackage{ae,aecompl}
\usepackage{bigints}

\usepackage{graphicx}	% Including figure files
\usepackage{amsmath}	% Advanced maths commands
\usepackage{amsmath, epsfig,natbib}
\usepackage{multicol}
\usepackage{color,ulem}
\usepackage{newtxtext,newtxmath}
\definecolor{webgreen}{rgb}{0,.5,0}
\definecolor{webbrown}{rgb}{.6,0,0}
\usepackage{subfigure}

\newcommand{\pc}{\>{\rm pc}}
\newcommand{\kpc}{\mbox{$\>{\rm kpc}$}}

\newcommand{\Gyr}{\mbox{$\>{\rm Gyr}$}}

\newcommand\degrees{^\circ}

\title [bar-induced dark gaps in disc galaxies] 
{Closing the gap: secular evolution of bar-induced dark gaps in presence of thick discs}
\author[Ghosh et al.]
	{Soumavo Ghosh,$^{1}$\thanks{E-mail: ghosh@mpia-hd.mpg.de}
     Dimitri A. Gadotti, $^{2}$
     Francesca Fragkoudi, $^{3}$ 
     Vighnesh Nagpal, $^{4}$ 
     Paola Di Matteo, $^{5}$ and
     \newauthor
     Virginia Cuomo $^{6}$\\
$^1$ Max-Planck-Institut f\"{u}r Astronomie, K\"{o}nigstuhl 17, D-69117 Heidelberg, Germany\\
$^2$ Centre for Extragalactic Astronomy, Department of Physics, Durham University, South Road, Durham DH1 3LE, UK\\
$^3$ Institute for Computational Cosmology, Department of Physics, Durham University, South Road, Durham DH1 3LE, UK\\
$^4$ Department of Astronomy, University of California, Berkeley, CA 94720, USA\\
$^5$ GEPI, Observatoire de Paris, PSL Research University, CNRS, Place Jules Janssen, 92195 Meudon, France\\
$^6$ Instituto de Astronomía y Ciencias Planetarias, Universidad de Atacama, Avenida Copayapu 485, 1530000 Copiapó, Atacama, Chile\\
}

 \date{Accepted 2024 July 20. Received 2024 July 02; in original form 2024 April 22}

\pubyear{2024}

\begin{document}
\label{firstpage}
\pagerange{\pageref{firstpage}--\pageref{lastpage}}
\maketitle
%&&&&&&&&&&&&&&&&&&&&&&&&&&&&&&&&&&&&&&
\begin{abstract} 
The presence of dark gaps, a preferential light deficit along the bar minor axis, is observationally well known. The properties of dark gaps are thought to be associated with the properties of bars, and their spatial locations are often associated with bar resonances. However, a systematic study, testing the robustness and universality of these assumptions, is still largely missing.  Here, we investigate the formation and evolution of bar-induced dark gaps using a suite of $N$-body models of (kinematically cold) thin and (kinematically hot) thick discs with varying thick disc mass fraction, and different thin-to-thick disc geometry. We find that dark gaps are a natural consequence of the trapping of disc stars by the bar. The properties of dark gaps (such as strength and extent) are well correlated with the properties of bars. For stronger dark gaps, the fractional mass loss along the bar minor axis can reach up to $\sim 60-80$ percent of the initial mass contained, which is redistributed within the bar. These trends hold true irrespective of the mass fraction in the thick disc and the assumed disc geometry.
In all our models harbouring slow bars, none of the resonances (corotation, Inner Lindblad resonance, and 4:1 ultra-harmonic resonance) associated with the bar correspond to the location of dark gaps, thereby suggesting that the location of dark gaps is not a \textit{universal proxy} for these bar resonances, in contrast with earlier studies.

\end{abstract}
%&&&&&&&&&&&&&&&&&&&&&&&&&&&&&&&&&&&&&&

\begin{keywords}
{galaxies: evolution - galaxies: bar - galaxies: kinematics and dynamics - galaxies: structure -  methods: numerical}
\end{keywords}
%&&&&&&&&&&&&&&&&&&&&&&&&&&&&&&&&&

\section{Introduction}
\label{sec:Intro}
%&&&&&&&&&&&&&&&&&&&&&&&&&&&&&&&&&

A substantial fraction of disc galaxies in the local Universe harbours a stellar bar in the central region. The bar fraction reaches up to $\sim 50$ percent in optical wavelengths while in infrared wavelengths, this bar fraction increases to around two-thirds of the whole disc galaxy population in the local Universe \citep[e.g. see][]{Eskridgeetal2000,Menedesetal2007,NairandAbraham2010,Mastersetal2011,Butaetal2015,Kruketal2017}.  Past observational studies revealed that the bar fraction as well as the bar properties vary with stellar mass and Hubble type \citep[e.g. see][]{Kormdendy1979,Aguerrietal2005,MarinovaandJogee2007,Gadotti2011,Aguerrietal2009,Butaetal2010,NairandAbraham2010,Barwayetal2011,Erwin2018}. High redshift ($z \sim 1$) disc galaxies host prominent bars as well with the bar fraction decreasing with redshift \citep[e.g. see][but also see \citet{Elmetal2004,Jogeeetal2004}]{Shethetal2008,Melvinetal2014,Simmonsetal2014}. Recent JWST observations further revealed the presence of conspicuous bars even at higher redshifts \citep[$z \sim 3$; ][]{Guoetal2022,LeConteetal2023,Costantinetal2023,Smailetal2023,Tsukui2023}.
 Several numerical studies demonstrated that in an $N$-body model, a bar often forms quite spontaneously \citep[e.g see][]{CombesandSanders1981,SellwoodandWilkinson1993,DebattistaSellwood2000,Athanassoula2003}. Furthermore, cosmological simulations showed that bar formation already starts at $z \sim 1$ or more \citep[e.g. see][]{Kraljicetal2012,Fragkoudietal2020,Fragkoudietal2021,Rosas-Guevaraetal2022,Fragkoudietal2024}. At these redshifts, the discs are known to be kinematically hot (and turbulent), more gas rich, and possess a massive thick disc. However, recent $N$-body simulations (with both thin and thick discs) demonstrated that even in presence of a massive thick disc (analogous to those high redshift galaxies), bars and boxy/peanut bulges can form, purely from the internal gravitational instability \citep[see][]{Ghoshetal2023a,Ghoshetal2023b}. 
\par
Past theoretical works demonstrated that as a bar grows over time, it continuously traps stars that are on nearly circular orbits onto the $x_1$ orbits that serves as a backbone for the bar structure \citep[e.g. see][]{ContopoulosandGrosbol1989,Athanassoula2003,BinneyTremaine2008}. This transforms an initial azimuthally smooth light profile into a rather radially bright light profile; thereby causing a light deficit or `dark gap' along the bar minor axis. The presence of dark gaps has been shown observationally in barred galaxies \citep[e.g. see][]{GadottianddeSouza2003,Kimetal2016,Buta2017}. Using a sample of barred galaxies from the \textit{Spitzer} Survey of Stellar Structure in Galaxies (S$^4$G), \citet{Kimetal2016} showed that the strength of the dark gap is strongly related to the bar size and to  bar-to-total light ratio, and the light deficit (along the bar minor axis) is directly proportional to bar size. In addition, \citet{Aguerrietal2023} showed that for about 90 percent of their chosen sample of barred galaxies from the MaNGA survey, the ratio of bar length to dark gap length remains greater than 1.2. Past $N$-body models showed that indeed the formation of dark gaps is linked with the growth of a stellar bar \citep[e.g. see][]{Kimetal2016,GhoshDiMatteo2023} and dark gaps  are more prominent and are located at larger radii as the bar evolves with time \citep{Aguerrietal2023,GhoshDiMatteo2023}. Past studies have associated the location of the dark gap with different resonances of the bar.  \citet{Buta2017}, using $\sim 50$ early-to-intermediate-type barred galaxies, associated the location of the dark gaps with the corotation of the bar.  On the other hand, recent studies by \citet{Krishnaraoetal2022} and \citet{Aguerrietal2023}, using a sample of MaNGA barred galaxies (and supplemented by an $N$-body model of a barred galaxy), showed that the locations of the dark gaps are associated with the 4:1 ultra-harmonic resonance of the bar. However, a systematic study of the variation of properties of dark gaps with the properties of bars as well as testing the robustness and universality of the association of the location of dark gaps with different resonances (associated with the bar) is still missing. We aim to pursue this here.
\par
In this work, we carry out a systematic study of the formation and temporal evolution of bar-driven dark gaps as well as the detailed study of the correlation (if any) between the properties of the dark gaps and the bar. To achieve that, we make use of a suite of $N$-body models with (kinematically hot) thick and (kinematically cold) thin discs, mimicking the presence of thick disc in disc galaxies \citep[e.g. see][]{Pohlenetal2004,Yoachim2006,Comeronetal2016,Kasparovaetal2016,Comeronetal2019,Pinnaetal2019b,Pinnaetal2019a,matigetal2021,Scottetal2021}. Within the suite of $N$-body models, we vary the thick disc mass fraction as well as consider different geometric configurations (varying ratio of the thin and thick disc scale lengths). One of these models were studied in context of properties of bars and boxy/peanut bulges \citep{Fragkoudietal2017} and the whole suite of models were studied concerning the formation of bars and boxy/peanut bulges in presence of thick discs  \citep{Ghoshetal2023a,Ghoshetal2023b}. Furthermore, as shown later in this work, the thin+thick models harbour a \textit{slow bar}, that is, with $R_{\rm CR}/ R_{\rm bar} > 1.4$ where $R_{\rm CR}$ is the location of bar corotation and $R_{\rm bar}$ is the bar length \citep[for further details, see][]{DebattistaSellwood2000}. Therefore, it is well-suited to perform a systematic investigation of formation and evolution of bar-induced dark gaps as well as studying the correlation of their properties with bar properties. In addition, we present here a detailed investigation of the robustness and universality of association of dark gaps with different bar resonances which is largely missing in the literature.
\par
The rest of the paper is organized as follows.
Section~\ref{sec:sim_setup} provides a brief description of the suite of $N$-body models used in this work. Section~\ref{sec:darklane_properties} presents our findings on the bar-induced dark gaps, their properties and the temporal evolution as well as the correlation between the properties of the dark gaps and the bar. Section~\ref{sec:mass_loss_growth} presents the details of mass re-distribution (within the bar region) as dark gaps grow over time. Section~\ref{sec:location_resonance} contains the results of the robustness and universality of of association of dark gaps with different bar resonances. Section \ref{sec:summary} summarizes the main findings of this work.

%&&&&&&&&&&&&&&&&&&&&&&&

\section{Simulation set-up \& $N$-body models}
\label{sec:sim_setup}
%&&&&&&&&&&&&&&&&&&&&&&&&&&&&&&&&&

%
%================================
\begin{table}
\centering
\caption{Key structural parameters for the equilibrium models.}
\begin{tabular}{ccccc}
\hline
\hline
 Model$^{(1)}$ & $f_{\rm thick}$$^{(2)}$ & $R_{\rm d, thin}$$^{(3)}$ & $R_{\rm d, thick}$$^{(4)}$ \\
\\
 && (kpc) & (kpc) \\
\hline
rthickS0.1 & 0.1 & 4.7 &  2.3  \\
rthickE0.1 & 0.1 & 4.7 &  4.7 \\
rthickG0.1 & 0.1 & 4.7 &  5.6  \\
rthickS0.3 & 0.3 & 4.7 &  2.3  \\
rthickE0.3 & 0.3 & 4.7 &  4.7  \\
rthickG0.3 & 0.3 & 4.7 &  5.6  \\
rthickS0.5 & 0.5 & 4.7 &  2.3  \\
rthickE0.5 & 0.5 & 4.7 &  4.7 \\
rthickG0.5 & 0.5 & 4.7 &  5.6 \\
rthickS0.7 & 0.7 & 4.7 &  2.3 \\
rthickE0.7 & 0.7 & 4.7 &  4.7 \\
rthickG0.7 & 0.7 & 4.7 &  5.6 \\
rthickS0.9 & 0.9 & 4.7 &  2.3  \\
rthickE0.9 & 0.9 & 4.7 &  4.7  \\
rthickG0.9 & 0.9 & 4.7 &  5.6   \\
\hline
\end{tabular}
%\centering
\newline{
(1) Name of the model; (2) thick disc mass fraction; (3) scale length of the thin disc; (4) scale length of the thick disc.}
\label{table:key_param}
\end{table}
%%
%================================

To carry out a systematic study of the properties and the temporal evolution of dark gaps, we make use of a suite of $N$-body models, each consisting of a thin and a thick stellar disc, and the whole system is embedded in a live dark matter halo. One such model is already presented in \citet{Fragkoudietal2017}.
In addition, these models have been thoroughly studied in recent works of \citet{Ghoshetal2023a} and \citet{Ghoshetal2023b} in connection with bar and boxy/peanut formation scenario under varying thick disc mass fractions. Here, we use a sub-sample (15 out of a total of 25 models) of the entire suite of thin+thick models to investigate bar-driven dark gaps and their temporal evolution with varying thick disc mass fraction.

\par
The details of the initial equilibrium models are already provided in \citet{Fragkoudietal2017} and \citet{Ghoshetal2023a}. For the sake of completeness, here we briefly mention the equilibrium models. Each of the thin and thick discs is modelled with a Miyamoto-Nagai profile \citep{MiyamatoandNagai1975}, having $R_{\rm d}$, $z_{\rm d}$, and $M_{\rm d}$ as the characteristic disc scale length, the scale height, and the total mass of the disc, respectively. The total stellar mass (thin and thick) is fixed to $1 \times 10^{11} M_{\odot}$ for all the models considered here while the fraction of stellar mass in the thick disc population ($f_{\rm thick}$) varies from 0.1 to 0.9.
The scale heights of the thin and thick discs are fixed to $0.3 \kpc$ and $0.9 \kpc$, respectively. The dark matter halo is modelled by a Plummer sphere \citep{Plummer1911}, having $R_{\rm H}$ ($= 10 \kpc$) and $M_{\rm dm}$ ($= 1.6 \times 10^{11} M_{\odot}$) as the characteristic scale length and the total halo mass, respectively. The dark matter halo parameters are kept fixed across the suite of thin+thick models considered here. The values of the key structural parameters for the thin and thick discs are mentioned in Table~\ref{table:key_param}. For this work, we analysed a total of 15 $N$-body models of such thin+thick discs.
\par
A total of $1 \times 10^6$ particles are used to model the stellar (thin+thick) disc while a total of $5 \times 10^5$ particles are used to model  the dark matter halo. The initial conditions of the discs are obtained using the iterative method algorithm \citep[for details, see][]{Rodionovetal2009}. For this work, we only constrained the density profile of the stellar discs while allowing the velocity dispersions (specifically the radial and vertical components) to vary in such a way that the system converged to an equilibrium solution. The corresponding radial profiles of velocity dispersion are shown in \citet[see their Fig. 1]{Fragkoudietal2017}. For further details, the reader is referred to \citet{Fragkoudietal2017} and \citet{Ghoshetal2023a}. The simulations are run using a TreeSPH code by \citet{SemelinandCombes2002}. A hierarchical tree method \citep{BarnesandHut1986} with an opening angle $\theta = 0.7$ is used for calculating the gravitational force which includes terms up to the quadrupole order in the multipole expansion. A Plummer potential is employed for softening the gravitational forces with a softening length $\epsilon = 150 \pc$. We evolved all the models for a total time of $9 \Gyr$.

\par

Within the suite of thin+thick disc models, we considered three different scenarios for the scale lengths of the two disc (thin and thick) components. In rthickE models, $R_{\rm d, thick} = R_{\rm d, thin}$;  in rthickS models, $R_{\rm d, thick} < R_{\rm d, thin}$; and in rthickG models, $R_{\rm d, thick} > R_{\rm d, thin}$ where $R_{\rm d, thin}$ and $R_{\rm d, thick}$ denote the scale length for the thin and thick disc, respectively. Following \citet{Ghoshetal2023a}, any thin+thick model is referred as a unique string `{\sc [model configuration][thick disc fraction]'}. {\sc [model configuration]} denotes the corresponding thin-to-thick disc scale length configuration, that is, rthickG, rthickE, or {rthickS whereas {\sc [thick disc fraction]} denotes the fraction of the total disc stars that are in the thick disc population (or equivalently, the mass fraction in the thick disc as all the disc particles have same mass).

\section{Properties of dark gaps and their correlation with bar properties}
\label{sec:darklane_properties}
%&&&&&&&&&&&&&&&&&&&&&&&&&&&&&&&&&&&&

Fig.~\ref{fig:darkgap_example} (left panel) shows the face-on surface brightness distribution (in mag arcsec$^{-2}$) for the model rthickS0.1, calculated at the end of simulation run ($t = 9 \Gyr$). We used a magnitude zero-point ($m_0$) of $22.5$ mag arcsec$^{-2}$ to create the surface brightness map from the intrinsic particle distribution. The same magnitude zero-point is used throughout this work. In addition, a conversion of $1 \kpc = 1$ arcsec is used throughout this work. This would place the mock galaxies (produced from the thin+thick models) at a redshift $z \sim 0.05$ with an assumed $\Lambda$CDM
cosmology with parameters $\Omega_{m} = 0.315$, $H_0 = 67.4$ km s$^{-1}$ Mpc$^{-1}$ \citep{Planck2020}. Moreover, we assumed a mass-to-light ratio ($\Upsilon$) in order to convert the mass distribution to the light distribution. Since our models include both the thin and thick disc stars, therefore a reasonable choice for the $\Upsilon_{T}/\Upsilon_{t}$ is required where $\Upsilon_T$ and $\Upsilon_t$ denote the assumed mass-to-light ratio for the thick and thin disc stars, respectively. Following \citet{Comeronetal2011}, we assumed three values of $\Upsilon_{T}/\Upsilon_{t}$ in this work, namely,  $\Upsilon_{T}/\Upsilon_{t}$ = 1, 1.2, and 2.4 \citep[for details see][]{Comeronetal2011}, and further checked how these assumed values of  $\Upsilon_{T}/\Upsilon_{t}$ affect the results concerning the strength and extent of dark gaps.  
Even a mere visual inspection of Fig.~\ref{fig:darkgap_example} reveals the presence of conspicuous dark gap, along the bar minor axis, for the model rthickS0.1. In Appendix~\ref{appen_correlation_darkgapLengthStrength_allmodels}, we show the face-on surface brightness distribution (in mag arcsec$^{-2}$), calculated at $t = 9 \Gyr$, for all thin+thick models considered here (see Fig.~\ref{fig:densmap_endsteps_appendix}). The presence of prominent dark gaps, for almost all the thin+thick models, is evident from Fig.~\ref{fig:densmap_endsteps_appendix}, similarly to the model rthickS0.1 shown here. In what follows, we quantify the strength and the extent of the dark gaps, and then investigate their temporal evolution with varying $f_{\rm thick}$ values.

%================================
% Begin figure
%================================
\begin{figure*}
\includegraphics[width=0.85\linewidth]{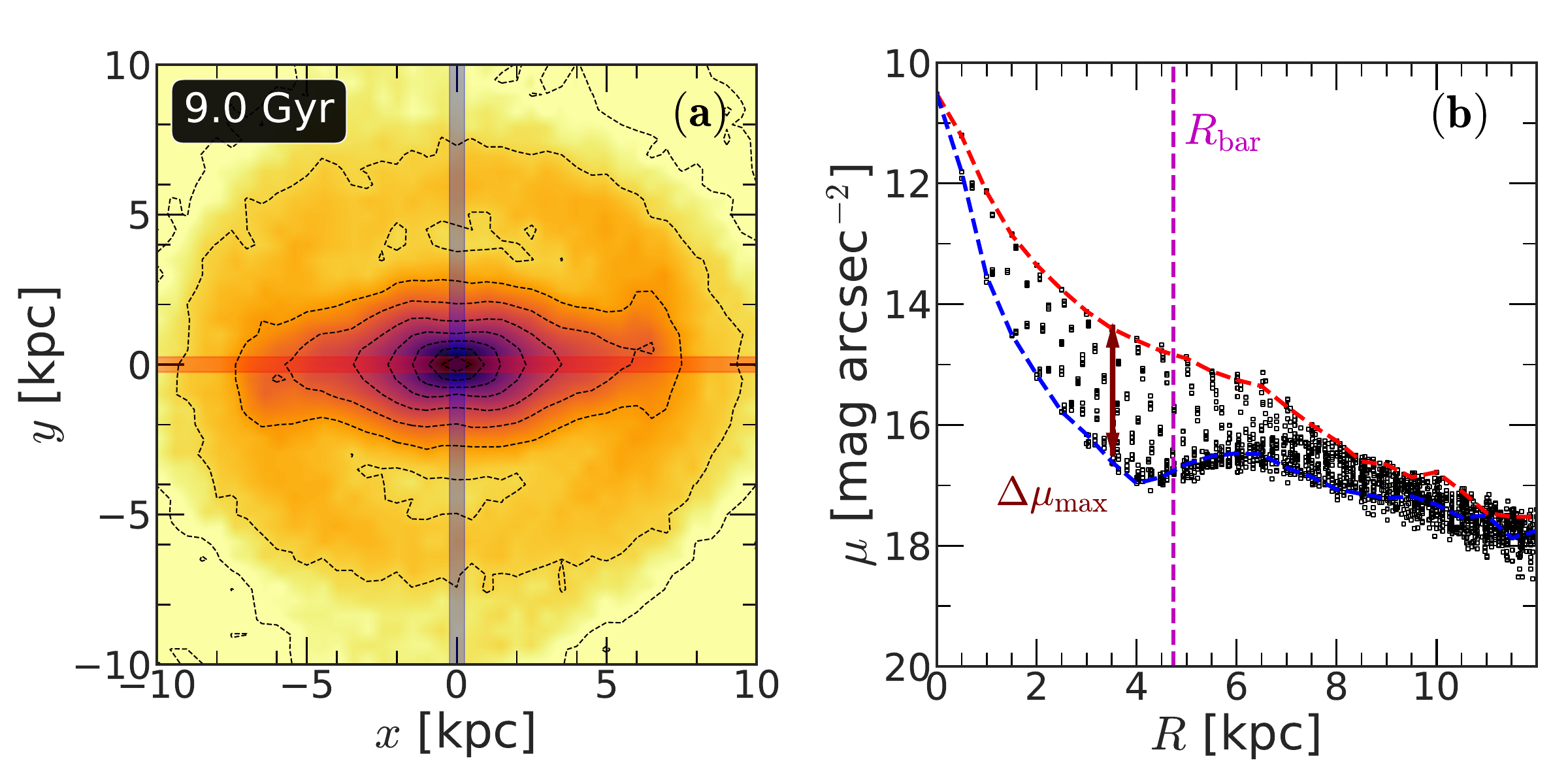}
\caption{\textit{Left panel:} Face-on surface brightness distribution, calculated using $\Upsilon_{T}/\Upsilon_{t} = 1.2$ (mass-to-light ratio for thick and thin disc stars) at $t = 9 \Gyr$ for the model rthickS0.1. The dashed black lines denote the contours of constant surface brightness. Here, a conversion of $1$ arcsec = $1 \kpc$ and a magnitude zero-point ($m_0$) of $22.5$ mag arcsec$^{-2}$ are used to create the surface brightness map from the intrinsic particle distribution. The red and blue lines denote the bar major and minor axis, respectively. \textit{Right panel:} Corresponding light profiles along the the bar major and minor axis (red and blue dashed lines, respectively). The radial location where the light deficit around the bar reaches its maximum ($\Delta \mu_{\rm max}$) is indicated by the maroon arrow. The vertical magenta line denotes the bar length, $R_{\rm bar}$. Each solid square represents a single pixel of the face-on surface brightness map.}
\label{fig:darkgap_example}
\end{figure*}
%================================
% End figure
%================================
%
\par
 To this aim, we first extract the surface brightness profiles along the bar major and minor axis while putting a slit of width, $\Delta = 0.5 \kpc$ in each direction. Following \citet{Kimetal2016}, at time $t$, we define the strength of the dark gap as the maximum light deficit, $\Delta \mu_{\rm max}$ between the bar major and minor axis. By definition, it is a non-negative quantity. In addition, we define the extent of the dark gap, $R_{\rm DG}$ where the maximum light deficit ($\Delta \mu_{\rm max}$) occurs, i.e., $ \Delta \mu (R = R_{\rm DG}) = \Delta \mu_{\rm max}$. An example of determining the $\Delta \mu_{\rm max}$ and $R_{\rm DG}$ is also shown in Fig.~\ref{fig:darkgap_example} (right panel) for the model rthickS0.1. 
%

%================================
% Begin figure
%================================
\begin{figure*}
\includegraphics[width=1.0\linewidth]{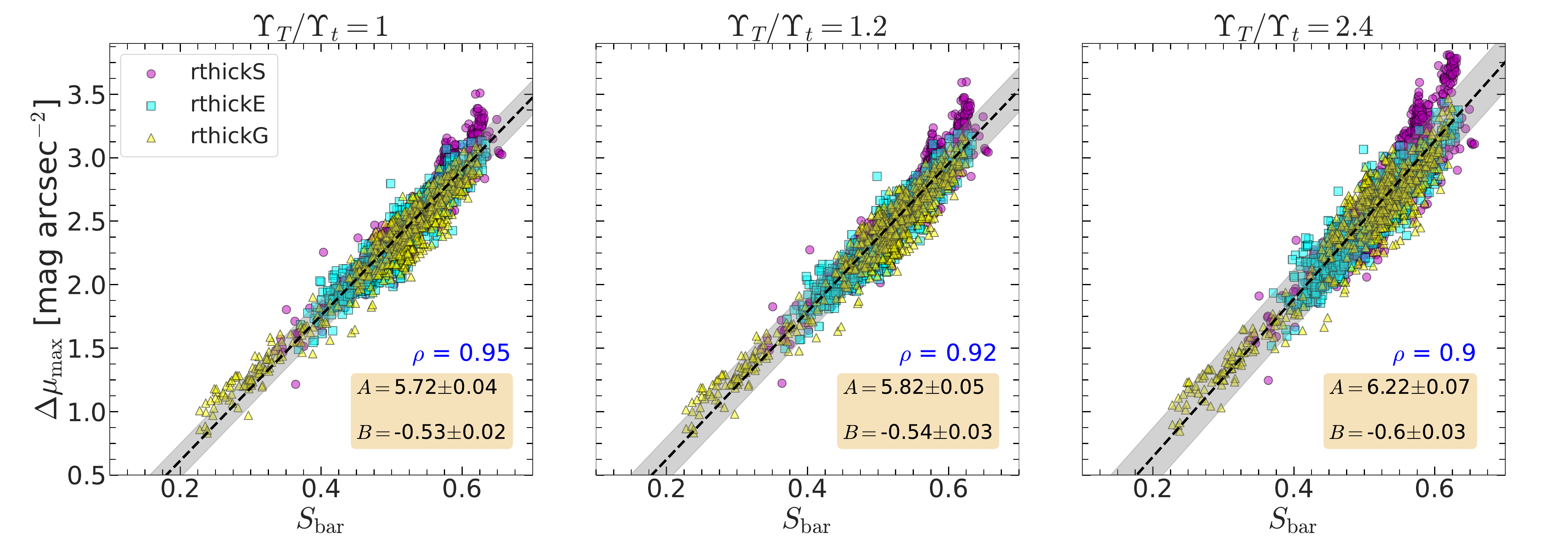}
\caption{Correlation between the bar strength, $S_{\rm bar}$ and the strength of the dark gap, $\Delta \mu_{\rm max}$, for all 15 thin+thick models considered here. \textit{Left panel} shows the correlation for $\Upsilon_{T}/\Upsilon_{t}$ =1 while the \textit{middle panel} and \textit{right panel} show the corresponding correlation for $\Upsilon_{T}/\Upsilon_{t}$ =1.2 and 2.4, respectively. The magenta circles denote the snapshots from the rthickS models, whereas the cyan squares and the yellow triangles denote the snapshots from the rthickE and rthickG models, respectively. The black dash line denotes the best-fit straight line (of the form $Y = AX+B$) while the grey shaded region denotes the 3-$\sigma$ scatter around the best-fit line. These two quantities remain strongly correlated (Pearson correlation coefficient, $\rho > 0.75$) throughout the entire temporal evolution. }
\label{fig:darklane_strength_correlation_consolidated}
\end{figure*}
%================================
% End figure
%================================

%
%================================
% Begin figure
%================================
\begin{figure*}
\includegraphics[width=0.925\linewidth]{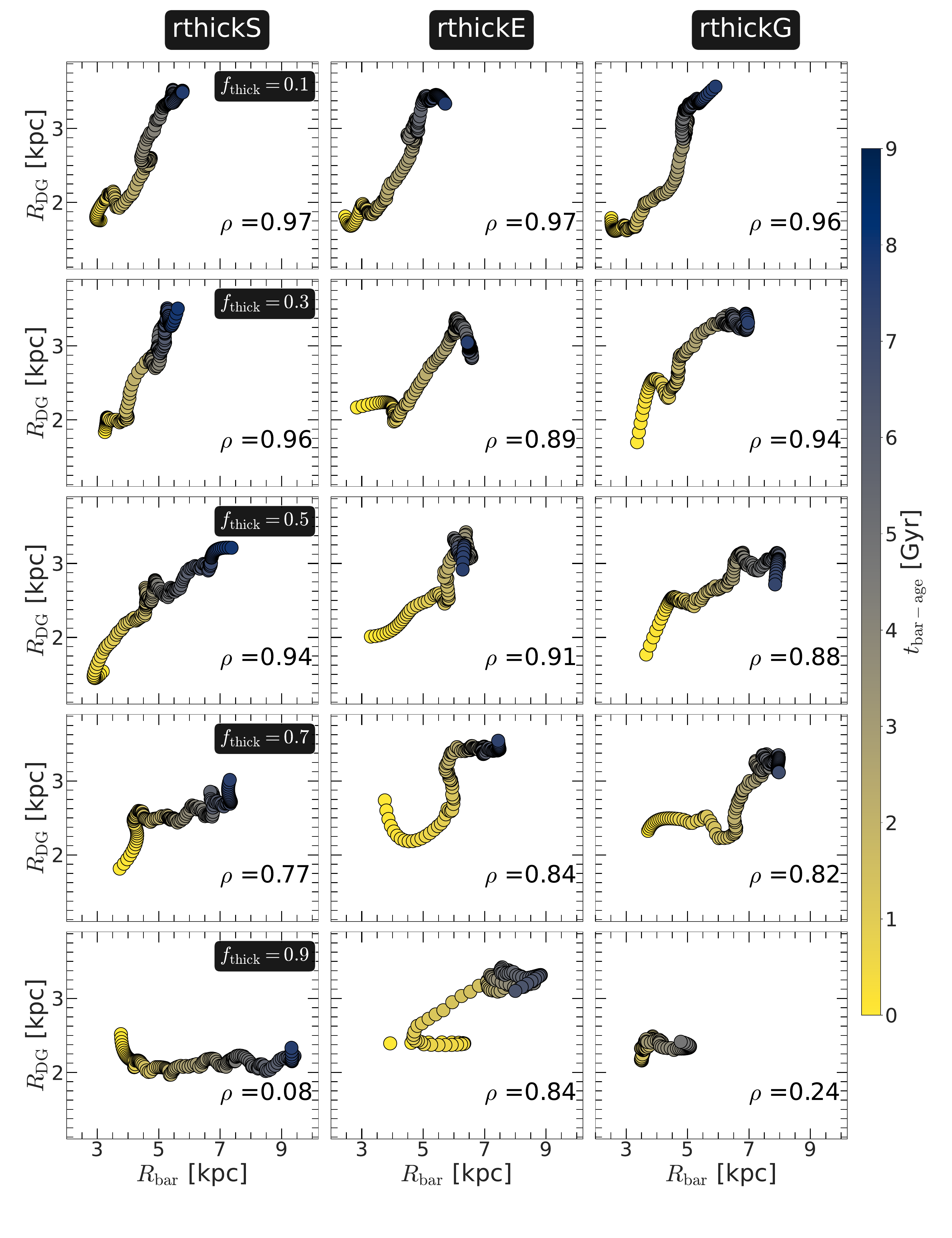}
\caption{Correlation between the bar length, $R_{\rm bar}$ and the extent of the dark gap, $R_{\rm DG}$, calculated for all thin+thick models, as a function of the bar age (see the colour bar). \textit{Left panels} correspond to the rthickS models whereas  \textit{middle panels} and \textit{right panels} correspond to the rthickE  and rthickG models, respectively. The thick disc fraction ($f_{\rm thick}$) varies from 0.1 to 0.9 (top to bottom), as indicated in the left-most panel of each row. In each case, the Pearson correlation coefficient, $\rho$ is calculated, and the corresponding value is quoted in each panel. These two quantities remain strongly correlated (Pearson correlation coefficient, $\rho > 0.75$) for all the thin+thick models, except for the models rthickS0.9 and rthickG0.9. For further details, see the text.}
\label{fig:bar_drakgap_length_correlation}
\end{figure*}
%================================
% End figure
%================================

%
%================================
% Begin figure
%================================
\begin{figure*}
\includegraphics[width=0.9\linewidth]{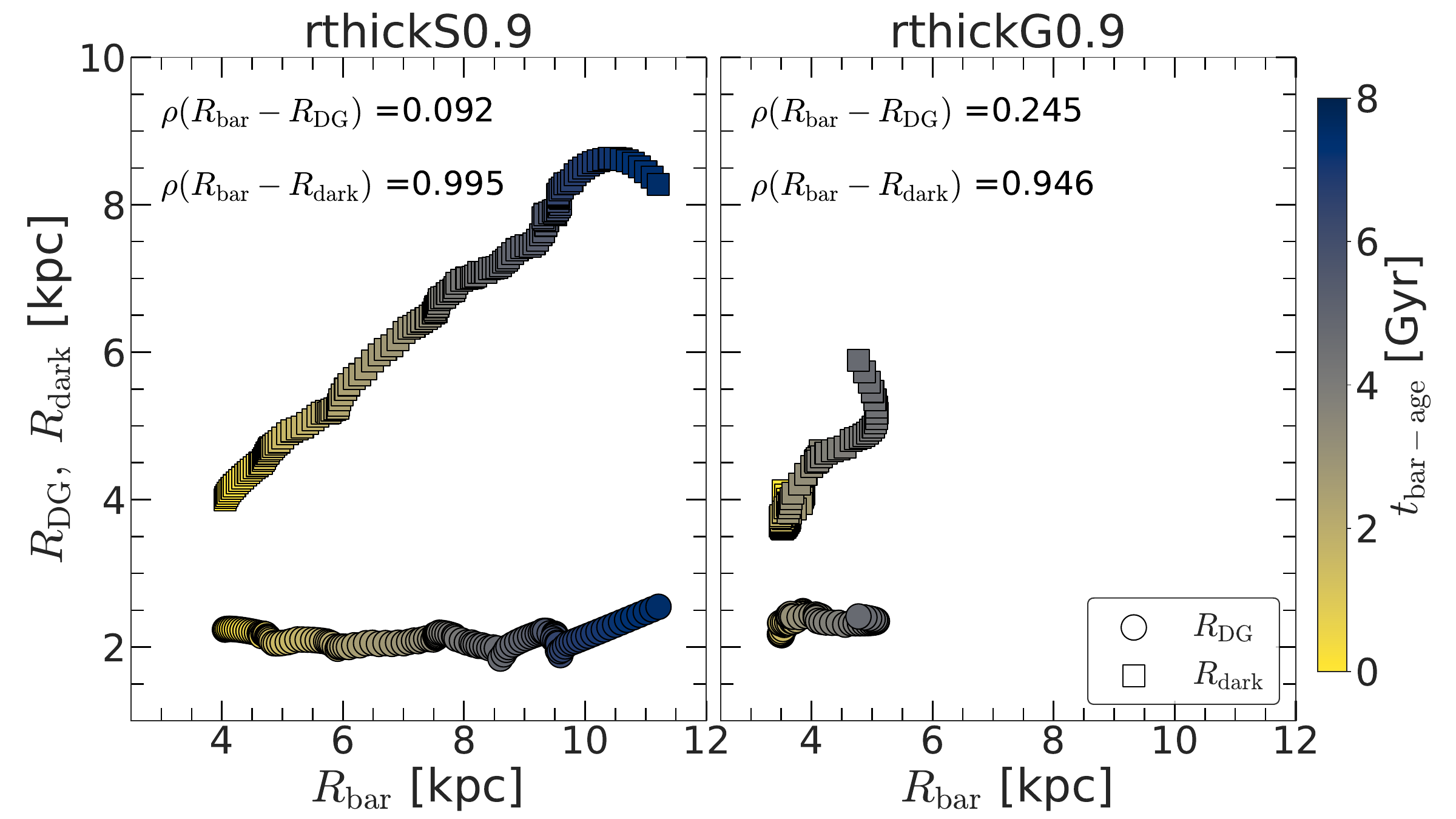}
\caption{Correlation between the bar length, $R_{\rm bar}$ and the extent of the dark gap, $R_{\rm DG}$ (in open circles), and between $R_{\rm bar}$ and $R_{\rm dark}$ (in open squares), calculated for the two thin+thick models, namely, rthickS0.9 (left panel) and rthickG0.9 (right panel), as a function of the bar age (see the colour bar). In each case, the Pearson correlation coefficient, $\rho$ is calculated, and the corresponding values are quoted. The two quantities, $R_{\rm bar}$ and $R_{\rm dark}$ remain strongly correlated (Pearson correlation coefficient, $\rho > 0.75$) for the models rthickS0.9 and rthickG0.9.}
\label{fig:bar_drakgap_length_comparison}
\end{figure*}
%================================
% End figure
%================================

 First, we investigate how the strength of the dark gaps is related to the bar strength. The bar strength, $S_{\rm bar}$, at time $t$, is defined as the maximum of the $m=2$ Fourier coefficient ($A_2/A_0$), that is $S_{\rm bar} (t) = max\{(A_2/A_0) (R, t)\}$, and these values are taken from \citet{Ghoshetal2023a}. The resulting correlation between the strengths of the dark gap and the bar, for the three assumed values of $\Upsilon_{T}/\Upsilon_{t}$, namely,  $\Upsilon_{T}/\Upsilon_{t}$ = 1, 1.2, and 2.4., for all thin+thick models considered here, are shown in Fig.~\ref{fig:darklane_strength_correlation_consolidated}. Note that in Fig.~\ref{fig:darklane_strength_correlation_consolidated} only the snapshots after the bar forms are considered for all thin+thick models.  Following \citet{Ghoshetal2023b}, we define the bar formation epoch $\tau_{\rm bar}$ as the epoch when the amplitude of the $m=2$ Fourier moment becomes greater than 0.2 and the corresponding phase angle, $\phi_2$ remains constant (within $3-5 \degrees$) within the extent of the bar. Therefore, the bar age is defined as $t_{\rm bar-age} = t - \tau_{\rm bar}$. Furthermore, we computed the Pearson correlation coefficient, $\rho$ to quantify the correlation. As seen clearly, when all the thin+thick models are taken together, the bar strength and the strength of the dark gaps remain strongly correlated ($\rho > 0.75$) for all three assumed values of $\Upsilon_{T}/\Upsilon_{t}$. This is not surprising since the bar strength is defined as a maximum of the $m=2$ Fourier coefficient of the density at a radial location $R$, and the dark gap strength is the (light-weighted and smoothed) peak-to-trough ratio of the density. Therefore, in limit where the density variation is sinusoidal with respect to azimuthal angle and the mass-to-light ratio is constant, these are perfectly correlated by construction. Fig.~\ref{fig:darklane_strength_correlation_consolidated} essentially demonstrates the fact that altering the mass-to-light ratio of the two components (thin and thick disc) does not significantly affect the fundamental conclusion that the strengths of bar and dark gap are inherently correlated.
\par
Next, we investigate how the extent of the dark gaps, $R_{\rm DG}$ evolves with time, and if there exists any correlation between the extent of the dark gap and the length of bar, in our thin+thick models. Following \citet{GhoshDiMatteo2023}, we define the bar length, $R_{\rm bar}$ as the radial extent where the amplitude of the $m=2$ Fourier moment ($A_2/A_0$)  drops to 70 percent of its peak value. We checked that  the ratio, $R_{\rm bar}/R_{\rm DG}$ always remains well above 1.2, at all times, for all thin+thick models. For the sake of brevity, they are not shown here. This finding is in agreement with \citet{Aguerrietal2023} who showed that for a majority (about 90 percent) of their sample of barred galaxies from the MaNGA survey, the ratio $R_{\rm bar}/R_{\rm DG}$ remains above 1.2.
\par
Lastly, we investigate if there exists any correlations between the bar length and the extent of the dark gap. This is shown in Fig.~\ref{fig:bar_drakgap_length_correlation} for all thin+thick models considered here. As seen from Fig.~\ref{fig:bar_drakgap_length_correlation}, the Pearson correlation coefficient $\rho$ remains well above 0.75 for almost all models; thereby indicating that the bar length and extent of dark gap are strongly correlated. However, for the models rthickS0.9 and rthickG0.9, the bar length and extent of dark gap are not correlated (see the corresponding $\rho$ values in Fig.~\ref{fig:bar_drakgap_length_correlation}).  We recall that $R_{\rm DG}$ is defined as the location where the maximum of peak-to-trough variation (along bar major and minor axes) occurs whereas $R_{\rm bar}$ is defined where the $A_2/A_0$ value drops to 70 percent of its peak value. Therefore, the question remains whether the temporal evolution of $R_{\rm DG}$ in these two models signifies a different evolutionary scenario for the dark gaps or is it due to the different definitions of $R_{\rm DG}$ and $R_{\rm bar}$ (one locating the peak while the other extends beyond the peak location). To verify that, we first calculated the radial profiles of $\Delta \mu$ for the three thin+thick models, namely, rthickS0.1, rthickS0.7, and rthickS0.9. This is shown in Appendix~\ref{appen_correlation_darkgapLengthStrength_allmodels} (see Fig.~\ref{fig:deltamu_radial_profiles_appendix} there).  As seen clearly from Fig.~\ref{fig:deltamu_radial_profiles_appendix}, the peak location of $\Delta \mu$ moves progressively towards the outer disc as the bar (and the dark gaps grow in strength) for the model rthickS0.1. However, for the model rthickS0.9, the peak location of $\Delta \mu$ does not move as much towards the outer disc region over time. This explains why $R_{\rm DG}$ values remain almost constant for the model rthickS0.9. The trend for the model rthickS0.7 falls somewhere in between these two above-mentioned trends. Lastly, to check whether a difference in defining $R_{\rm DG}$ and $R_{\rm bar}$ (one locating the peak while the other extends beyond the peak location) impacts the inference of correlation between the length of bars and dark gaps, we introduce a new metric to define the extent of the dark gap, namely, $R_{\rm dark}$ which is defined as the location where $\Delta \mu$ value falls to 70 percent of its peak value. The corresponding correlation between $R_{\rm bar}$ and $R_{\rm dark}$, computed for the models rthickS0.9 and rthickG0.9 are shown in Fig.~\ref{fig:bar_drakgap_length_comparison}. As seen clearly from Fig.~\ref{fig:bar_drakgap_length_comparison}, $R_{\rm bar}$ and $R_{\rm dark}$ remain strongly correlated ($\rho > 0.75$) for the models rthickS0.9 and rthickG0.9. We checked  that $R_{\rm bar}$ and $R_{\rm dark}$ remain strongly correlated for other thin+thick models as well. For the sake of brevity, we have not shown it here. This emphasises that when a uniform definition is used to define the bar length and the length of the dark gap, they remain strongly correlated over the entire evolutionary phase. A similar discrepancy exists in measuring the bar length from the peak location of the $A_2/A_0$ values and the location where the $A_2/A_0$ value drops to  the 70 percent of its peak value \citep[for further details, see][]{GhoshDiMatteo2023}. Our findings here outline the importance for a uniform definition for the bar and dark gap lengths, and demonstrate how different definitions for the bar and dark gap lengths might lead to an erroneous conclusion.
\par
To conclude, our systematic study demonstrates that the dark gap in a disc galaxy is essentially a part and parcel of the dynamical effect of a bar as the bar continuously redistributes the stars onto more radially elongated orbits; thereby producing a dearth of stellar density along the bar minor axis. The strength and the extent of these dark gaps can be used a robust proxy for the bar strength and length, respectively. 
This has a direct implication for the observational study of the dark gap and the bar properties. For a barred galaxy, observed at an intermediate inclination, the quantification of bar strength (calculated via the $m=2$ Fourier coefficient) can often be cumbersome, involving de-projection of the image (and associated uncertainty) and also critically depends of the resolution of the photometric image. As our findings demonstrate, the strength of the dark gaps, which are quite straightforward to compute for an observed galaxy, can serve as an excellent proxy for the bar strength. A similar argument applies for using the extent of the dark gaps as an excellent proxy for the bar length as well.

\section{Growth of dark gaps and the associated mass re-distribution}
\label{sec:mass_loss_growth}
%&&&&&&&&&&&&&&&&&&&&&&&&&&&&&&&&&

In the previous section, we demonstrated that the properties (strength and extent) of the dark gaps show a strongly correlated evolution with the bar properties (strength and length) in all thin+thick models considered here. As the bar grows in strength, it continuously traps more stars into more radially elongated orbits, thereby making an initial azimuthally uniform light profile into a rather radially bright light profile. Since we are dealing with $N$-body models here, and therefore, without any assumption of mass-to-light ($M/L$) ratio (as commonly done in observations), we can quantify the mass loss along the bar minor axis as the dark gaps grow with time. To quantify the fractional mass change in the $(x-y)$-plane (face-on configuration), at time $t$, we define

\begin{equation}
\delta M_{*} (x, y, t) = \frac{M_{*} (x, y, t) - M_{*} (x, y, t=0)}{M_{*} (x, y, t=0)}\,,
\label{eq:mass_loss_2D}
\end{equation}
\noindent where $M_{*} (x, y, t)$ denotes the stellar (thin+thick) mass at the spatial location $(x, y)$ at time $t$. The corresponding face-on  distribution of the fractional mass change, at different times (capturing different phases of bar evolution) is shown in Fig.~\ref{fig:fractional_massLoss_2d} for the model rthickS0.1. As seen clearly from Fig.~\ref{fig:fractional_massLoss_2d}, stellar mass gets enhanced along the bar major-axis, and simultaneously there is a continuous mass deficit along the bar minor axis. At times, when the bar (and hence, the dark gaps) is quite strong, the mass deficit along the bar minor axis can reach up to $\sim 60-80$ percent of its initial ($t =0$) mass.  

%================================
% Begin figure
%================================
\begin{figure*}
\includegraphics[width=0.9\linewidth]{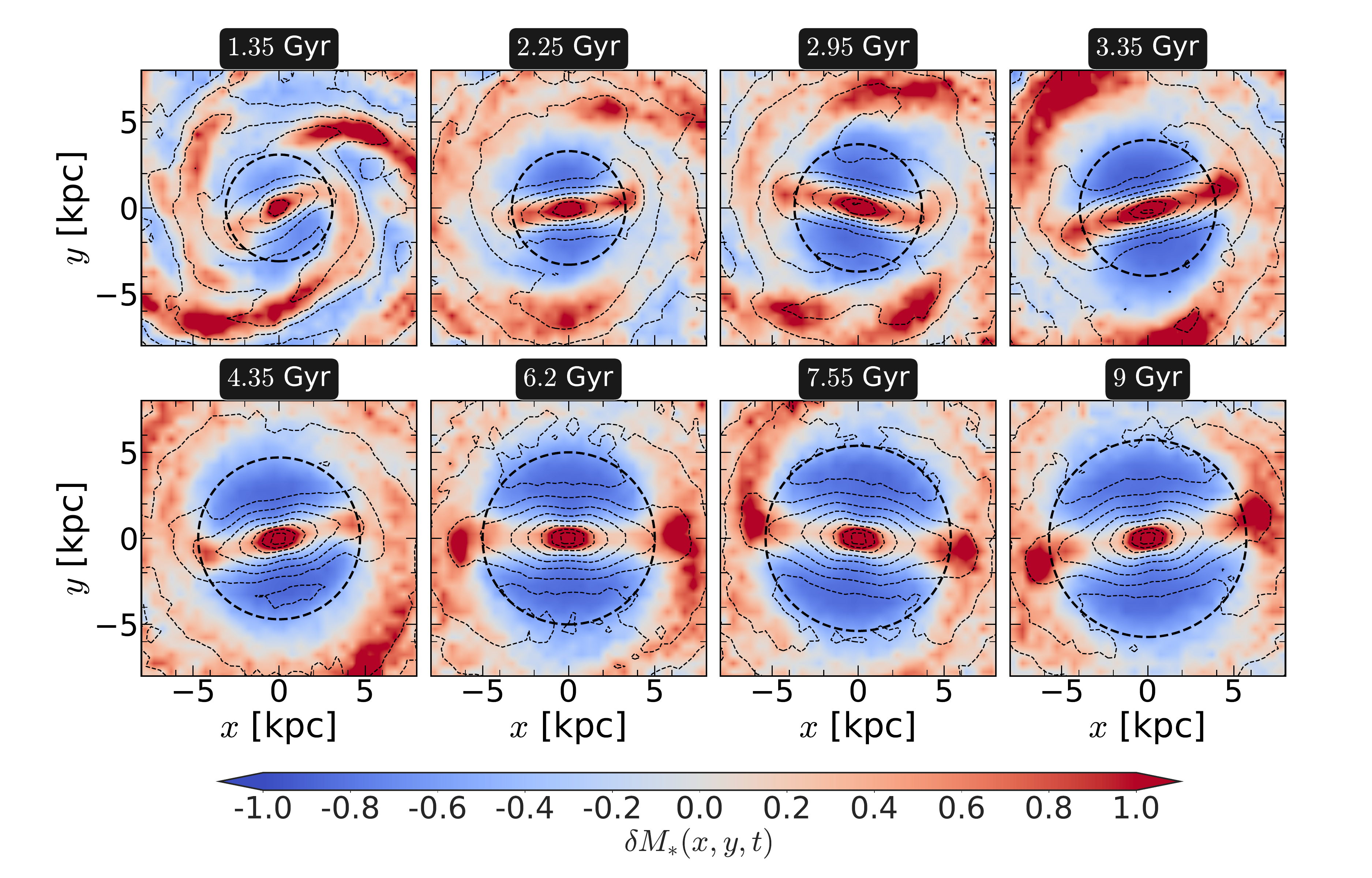}
\caption{Face-on distribution of the fractional mass change, $\delta M_{*} (x, y, t)$ (see Eq.~\ref{eq:mass_loss_2D}) at different times (capturing different phases of bar evolution) for the model rthickS0.1. A positive $\delta M_{*} (x, y, t)$ denotes the mass gain while a negative $\delta M_{*} (x, y, t)$ implies mass loss at a location $(x, y)$. The black dashed lines denote the contours constant surface density. The black circle denotes the bar length, $R_{\rm bar}$. As the bar evolves with time, stellar mass gets enhanced along the bar major axis (see the red regions), and simultaneously there is a continuous mass deficit along the bar minor axis (see the blue regions). }
\label{fig:fractional_massLoss_2d}
\end{figure*}
%================================
% End figure
%================================

Next, to quantify the fractional mass change, at time $t$, along the bar minor axis, we define

\begin{equation}
\delta M_* (y_{\rm minor}, t) = \frac{M_* (y_{\rm minor}, t) - M_* (y_{\rm minor}, t=0)}{M_* (y_{\rm minor}, t=0)}\,,
\label{eq:mass_loss_FranStyle}
\end{equation}

\noindent where $y_{\rm minor}$ denotes the spatial location along the bar minor axis, and $M_* (y_{\rm minor}, t)$ denotes the stellar mass at a spatial location $y_{\rm minor}$ along the bar minor axis at time $t$. A positive value of $\delta M_* (y_{\rm minor}, t)$ denotes mass increase whereas a negative value of $\delta M_* (y_{\rm minor}, t)$ denotes mass loss at a certain time $t$. The corresponding temporal evolution of $\delta M_* (y_{\rm minor}, t)$, calculated with $\Delta y_{\rm minor} = 1 \kpc$, for the thin+thick model rthickS0.1 is shown in Fig.~\ref{fig:mass_loss_along_minorAxis_FranStyle}. As seen from Fig.~\ref{fig:mass_loss_along_minorAxis_FranStyle}, spatial location corresponding to $y_{\rm minor} < 1 \kpc$ falls in the part of the bar structure, and it shows substantial mass increase (compare Figs. ~\ref{fig:fractional_massLoss_2d} and ~\ref{fig:mass_loss_along_minorAxis_FranStyle}). However,  spatial locations falling within the region of dark gap ($1 < y_{\rm minor}/ \kpc < 5$), show substantial mass loss (i.e. $\delta M_* (y_{\rm minor}, t) < 0$) over the course of the evolution. The temporal evolution of  $\delta M_* (y_{\rm minor}, t)$ beyond $y_{\rm minor} = 6 \kpc$ shows moderate increase at initial times, and this is due to the fact that over time, the disc grows in the outward direction. 
%
%================================
% Begin figure
%================================
\begin{figure}
\includegraphics[width=1.02\linewidth]{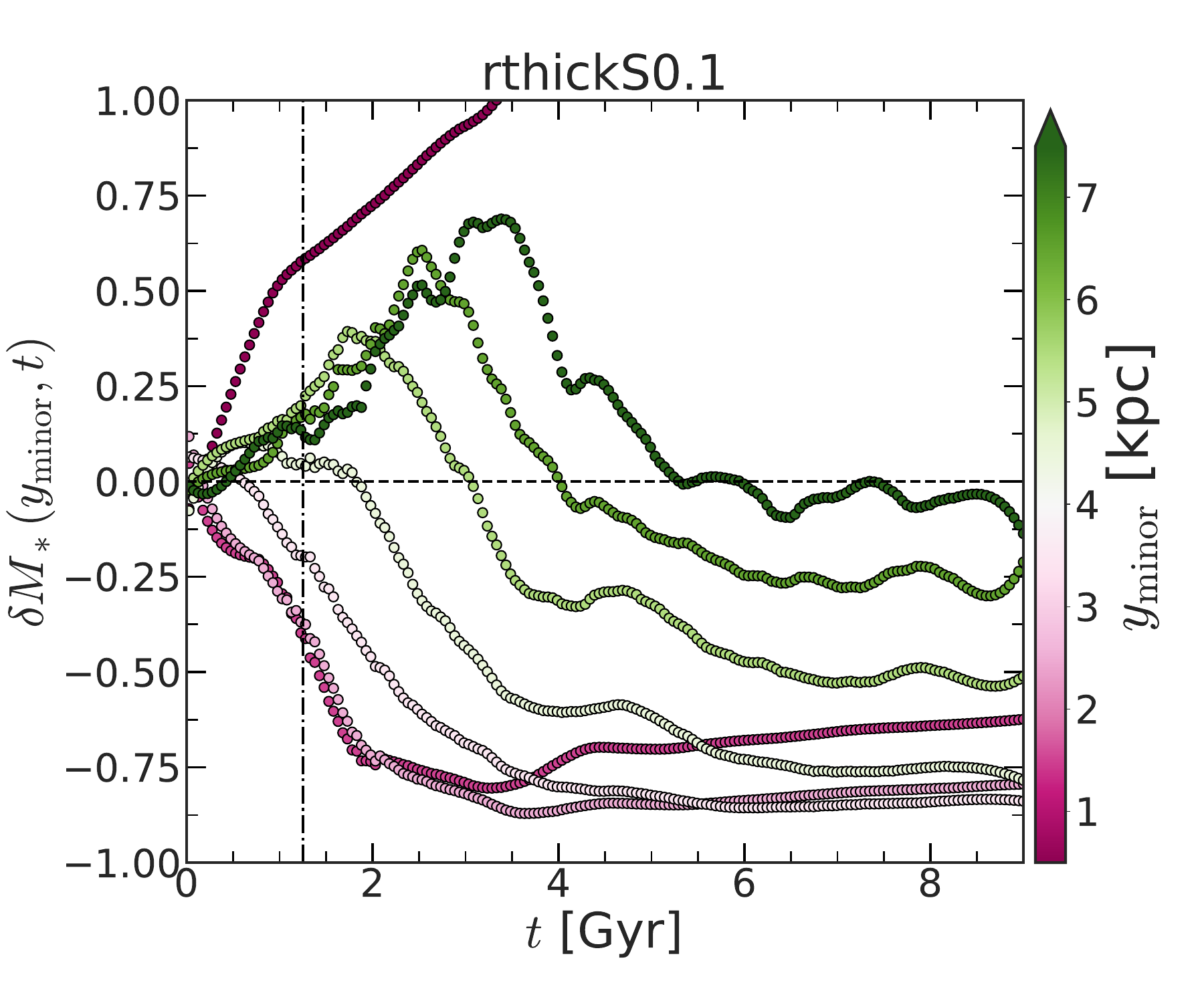}
\caption{Fractional mass loss, at different spatial locations along the bar minor axis, $\delta M_{*} (y_{\rm minor}, t)$ (Eq.~\ref{eq:mass_loss_FranStyle}), as a function of time for the thin+thick model rthickS0.1. The colour bar shows the spatial locations along the bar minor axis. The vertical dash-dotted line denotes the epoch of bar formation, $\tau_{\rm bar}$. $1 < y_{\rm minor}/ \kpc < 5$ denotes the region of the dark gap. For details, see section~\ref{sec:mass_loss_growth}.}
\label{fig:mass_loss_along_minorAxis_FranStyle}
\end{figure}
%================================
% End figure
%================================

%================================
% Begin figure
%================================
\begin{figure*}
\includegraphics[width=0.95 \linewidth]{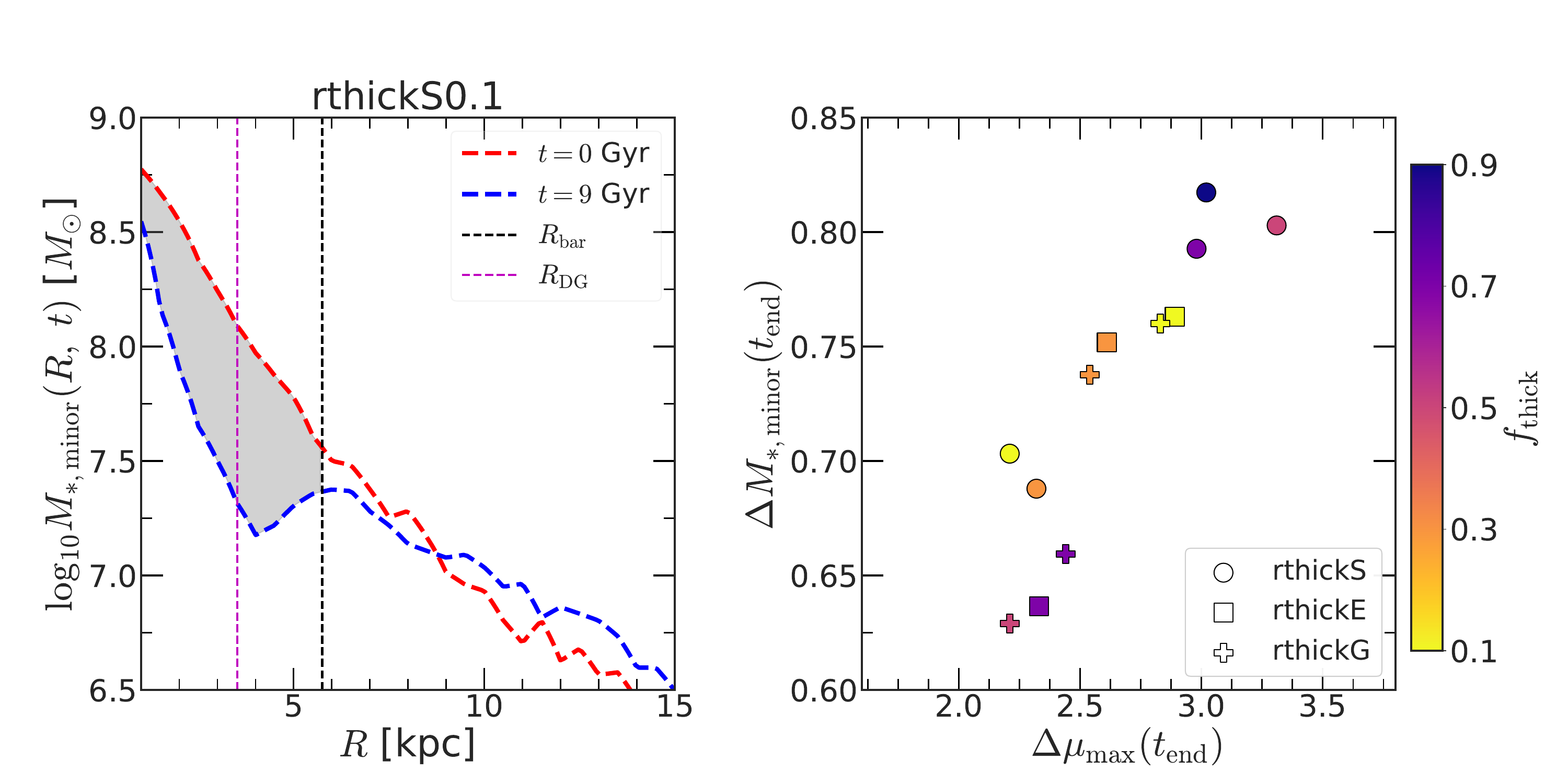}
\caption{\textit{Left panel}: Radial distribution of stellar (thin+thick) mass (in logarithmic scale), along the bar minor axis, at the beginning and at the end of the simulation run ($t_{\rm end} = 9 \Gyr$), for the model rthickS0.1. The vertical black dashed line denotes the bar extent ($R_{\rm bar}$) while vertical magenta dashed line denotes the extent of the dark gap ($R_{\rm DG}$) at $t = t_{\rm end}$. The grey shaded region denotes the fractional mass loss, $\Delta M_{*, \rm {minor}} (t_{\rm end})$ (see Eq.~\ref{eq:mass_loss_1D}), along the bar minor axis. \textit{Right panel}: correlation between the fractional mass loss along the bar minor axis ($\Delta M_{*, \rm {minor}} (t_{\rm end})$) and the strength of the dark gap $\Delta \mu_{\rm max} (t_{\rm end})$ for all thin+thick models considered here. The thick disc mass fraction ($f_{\rm thick}$) is shown in the colour bar. $\Upsilon_{T}/\Upsilon_{t}$ =1.2 is used to compute the values of $\Delta \mu_{\rm max}$.}
\label{fig:fractional_massLoss_1d}
\end{figure*}
%================================
% End figure
%================================

Lastly, to carry out a uniform comparison on fractional mass loss (within the extent of the bar) along the bar minor axis for all the thin+thick models considered here, we define

\begin{equation}
\Delta M_{*, \rm {minor}} (t_{\rm end}) =\frac{ \bigints_{R_{\rm in}}^{R_{\rm bar}} \left[ M_{*, \rm {minor}} (R, t_{\rm end}) - M_{*, \rm {minor}} (R, t =0) \right]  \ dR}{\bigints_{R_{\rm in}}^{R_{\rm bar}} M_{*, \rm {minor}} (R, t =0)  \ dR} \,,
\label{eq:mass_loss_1D}
\end{equation}
\noindent where we assumed $R_{\rm in} = 0.5 \kpc$ and $M_{*, \rm {minor}} (R, t)$ is the mass at a radial location $R$ at time $t$ along the bar minor axis. In Fig.~\ref{fig:fractional_massLoss_1d} (left panel), we show one such example of the fractional mass loss along the bar minor axis for the model rthickS0.1 (see the grey shaded region). Next, we compute the fractional mass loss (within the extent of the bar) along the bar minor axis using Eq.~\ref{eq:mass_loss_1D}, for all the thin+thick models considered here. This is shown in Fig.~\ref{fig:fractional_massLoss_1d} (right panel). The fractional mass loss (within the extent of the bar) along the bar minor axis, calculated at the end of the simulation run ($t_{\rm end} = 9 \Gyr$), is strongly correlated with the maximum light deficit, $\Delta \mu_{\rm max}$. For some thin+thick models showing stronger dark gaps (and harbouring stronger bar), the fractional mass loss (within the extent of the bar) along the bar minor axis can reach up to $\sim 80$ percent of the initial mass contained within the bar region (see right panel of Fig.~\ref{fig:fractional_massLoss_1d}). This strong correlation between the $\Delta M_{*, \rm {minor}} (t_{\rm end})$ and $\Delta \mu_{\rm max} (t_{\rm end})$ further supports the scenario of the growth of dark gaps as a result of continuous trapping of stars that are on nearly-circular orbits onto the more elongated orbits by the bar.

\section{Dark gaps and resonance locations}
\label{sec:location_resonance}
%&&&&&&&&&&&&&&&&&&&&&&&&&

%================================
% Begin figure
%================================
\begin{figure}
\includegraphics[width=0.98\linewidth]{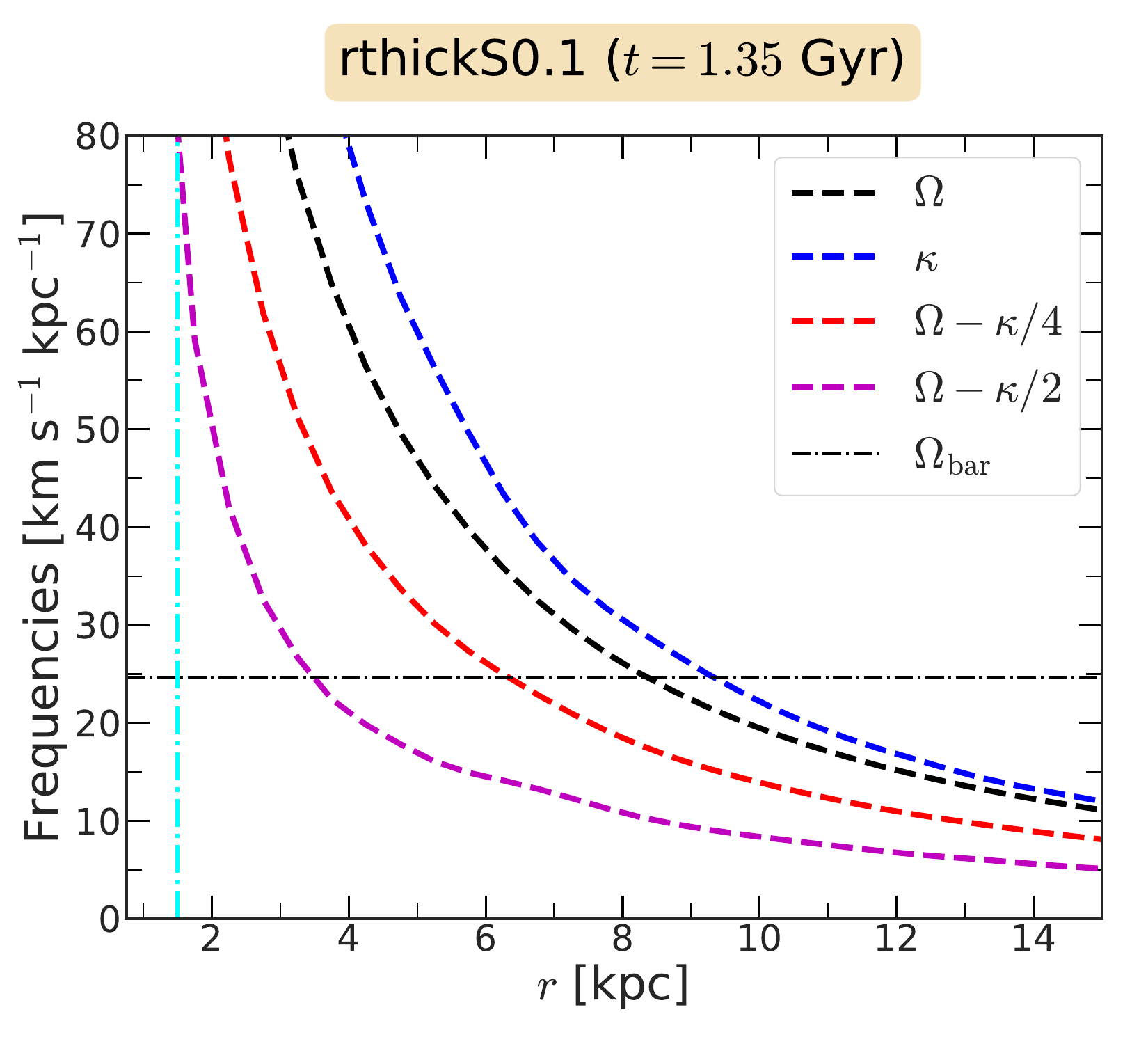}
\caption{Radial variation of the circular frequency ($\Omega$), epicyclic frequency ($\kappa$), $\Omega -\kappa/2$, and $\Omega- \kappa/4$ (see the legend), calculated at $t = 1.35 \Gyr$ for the model rthickS0.1. The horizontal dash-dotted black line denotes the bar pattern speed ($\Omega_{\rm bar}$) at that epoch while the vertical cyan line denotes the location of the dark gap ($R_{\rm DG}$) at the same epoch.}
\label{fig:circularVel_example}
\end{figure}
%================================
% End figure
%================================

%================================
% Begin figure
%================================
\begin{figure}
\includegraphics[width=1.02\linewidth]{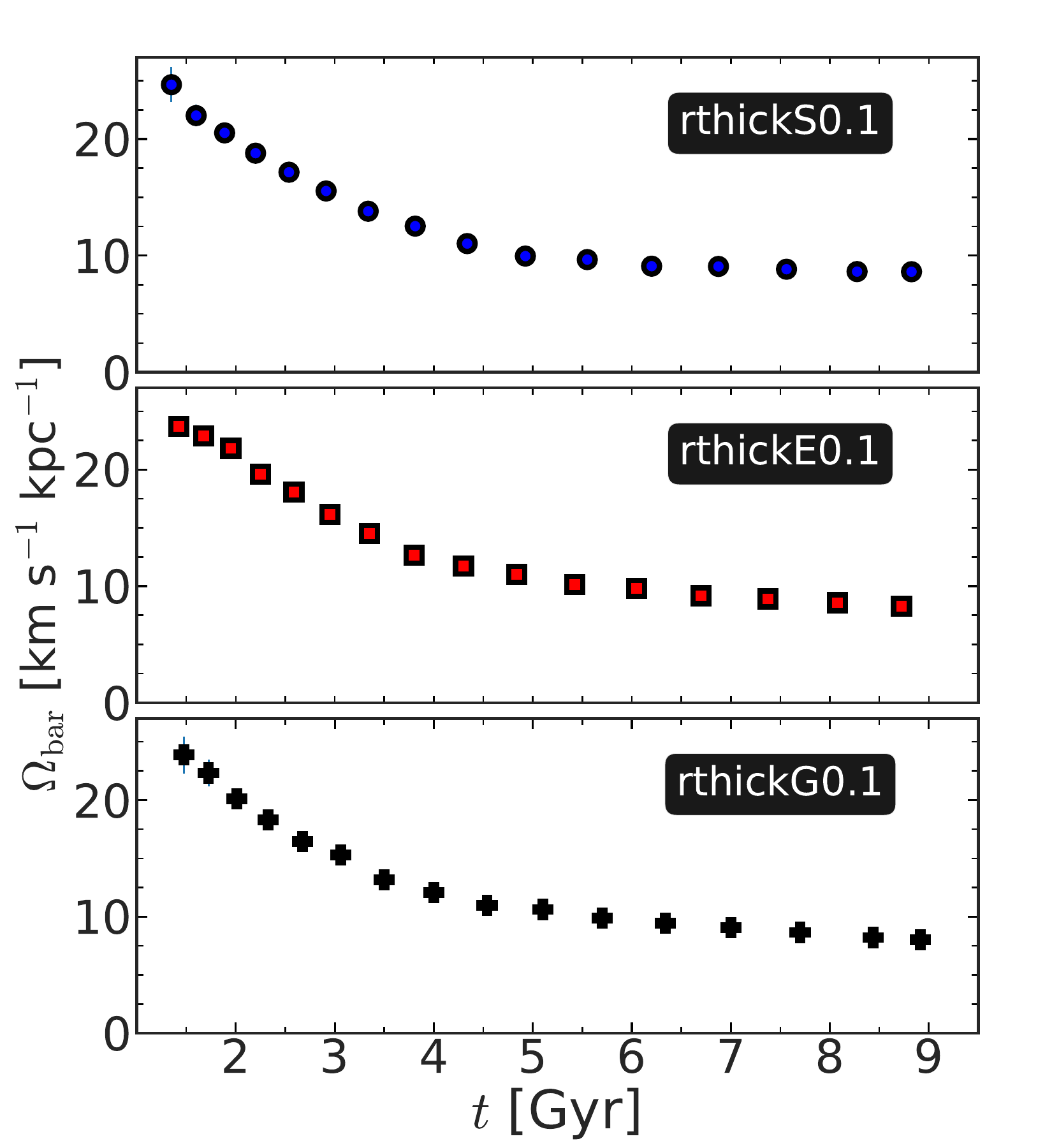}
\caption{Temporal evolution of the bar pattern speed ($\Omega_{\rm bar}$) for three thin+thick models. In each case, the bar pattern speed decreases substantially over time. }
\label{fig:pattern speed_example}
\end{figure}

%================================
% End figure
%================================

%================================
% Begin figure
%================================
\begin{figure*}
\includegraphics[width=0.88\linewidth]{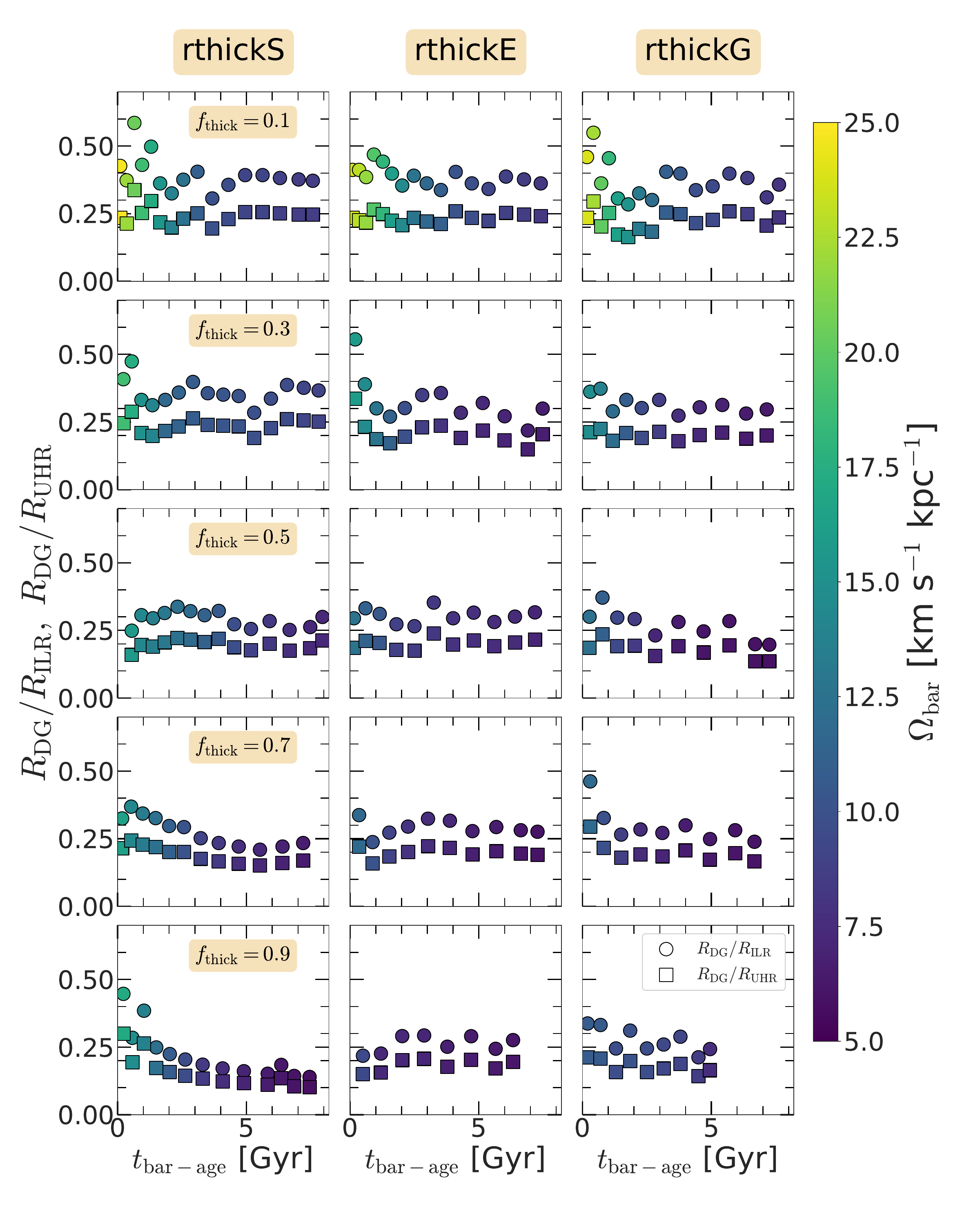}
\caption{Evolution of the ratios $R_{\rm DG}/R_{\rm ILR}$ (filled circles) and $R_{\rm DG}/R_{\rm UHR} $ (filled squares) with bar age ($t_{\rm bar-age}$), for all thin+thick models considered here. The points are colour-coded by the bar pattern speed ($\Omega_{\rm bar}$) values.
 \textit{Left panels} show for the rthickS models whereas  \textit{middle panels} and \textit{right panels} show for the rthickE  and rthickG models, respectively. The thick disc fraction ($f_{\rm thick}$) varies from 0.1 to 0.9 (top to bottom), as indicated in the left-most panel of each row. }
\label{fig:ilr_uhr_locations_allmodels}
\end{figure*}
%================================
% End figure
%================================

Past studies have associated the location of the dark gap ($R_{\rm DG}$) with different resonances of the bar (for details, see section~\ref{sec:Intro} and references therein). However, the robustness and universality of this trend has not been tested so far in the literature. We pursue it here. 
\par
In order to measure the correlation between the extent of the dark gaps and different resonances (associated with the bar), first we need to compute the circular velocity ($v_c$) and the bar pattern speed ($\Omega_{\rm bar}$) at different times for all thin+thick models considered here.  At time $t$, the circular velocity, $v_{\rm c}$ is calculated as
\begin{equation}
v_{\rm c} (R) = \frac{G M(\leq r)}{r} 
\label{eq:asy_drift1}
\end{equation}
\noindent Here, $M(\leq r)$ denotes the mass enclosed within a spherical radius $r$. Once we derive the circular velocity, the corresponding circular frequency $\Omega$ and the epicyclic frequency $\kappa$ are derived using $\Omega = v_{\rm c}/R$, and $\kappa^2 = 4 \Omega^2 + d \Omega^2/dR$ \citep[for details, see][]{BinneyTremaine2008}. The corresponding radial profiles of $\Omega$, $\kappa$, $\Omega -\kappa/2$, and $\Omega- \kappa/4$, calculated at $t = 1.35 \Gyr$ for the model rthickS0.1 is shown in Fig.~\ref{fig:circularVel_example}.
\par
In order to determine the location for the corotation, the 2:1 Inner Lindblad resonance, and the 4:1 ultra-harmonic resonance, we need to calculate the bar pattern speed. Following \citet{Ghoshetal2022} and \citet{GhoshDiMatteo2023}, we measure the bar pattern speed ($\Omega_{\rm bar}$) by fitting a straight line to the temporal variation of the phase-angle ($\phi_2$) of the $m=2$ Fourier mode. The underlying assumption is that the bar rotates rigidly with a single pattern speed in that time-interval. We follow this technique to compute the bar pattern speed ($\Omega_{\rm bar}$) as a function of time, for all thin+thick models considered here. The corresponding temporal evolution of bar pattern speed, for three such thin+thick models are shown in Fig.~\ref{fig:pattern speed_example}. The bar pattern speed decreases drastically during the entire evolutionary phase of the bar (see Fig.~\ref{fig:pattern speed_example}). The radial locations where the bar pattern speed, $\Omega_{\rm bar}$ intersects with $\Omega$, $\Omega - \kappa/2$, and $\Omega -\kappa/4$, determines the locations of the corotation, 2:1 Inner Lindblad resonance, and 4:1 ultra-harmonic resonance, respectively (see Fig.~\ref{fig:circularVel_example}). We checked that the ratio of $R_{\rm CR}$ (location of bar corotation) to bar length ($R_{\rm bar}$) always remains greater than 1.4, and this trend remains true for almost all thin+thick models considered here.Therefore, bars present in our thin+thick models qualify as \textit{slow bars}. The detailed study of the temporal evolution of $\Omega_{\rm bar}$ with varying thick disc mass fraction ($f_{\rm thick}$) is beyond the scope of this work and will be addressed in a future study.

 Using the rotation curve and the bar pattern speed, we calculated the locations of the corotation ($R_{\rm CR}$), the 2:1 Inner Lindblad resonance ($R_{\rm ILR}$), and the 4:1 ultra-harmonic resonance ($R_{\rm UHR}$) at different times, for all thin+thick models considered here (for an example, see Fig.~\ref{fig:circularVel_example}). We checked that the values of $R_{\rm CR}/ R_{\rm DG}$ remain well above unity for all thin+thick models considered here, thereby implying that the locations of the dark gaps are not associated with the corotation resonance of the bar in our bar models. This finding is similar to the conclusion drawn by \citet{Krishnaraoetal2022}. Furthermore, in Fig.~\ref{fig:ilr_uhr_locations_allmodels}, we show the temporal evolution of the quantities $R_{\rm DG}/R_{\rm ILR}$ and $R_{\rm DG}/R_{\rm UHR}$ as a function of bar age ($t_{\rm bar-age}$), for all thin+thick models considered here. The ratio of the $R_{\rm UHR}$ and the extent of dark-daps, $R_{\rm DG}$, remains almost constant (especially at later bar evolutionary phases) and this holds true for almost all thin+thick models considered here. However, the ratio $R_{\rm DG}/R_{\rm UHR}$ remains well below unity, for all the models, thereby demonstrating that the dark gaps are also \textit{not} associated with the location of 4:1 ultra-harmonic resonances for any of the models considered here, in contrast with the results presented in \citet{Krishnaraoetal2022} and \citet{Aguerrietal2023}. In addition, the temporal evolution of the ratio $R_{\rm DG}/R_{\rm ILR}$ show somewhat oscillatory behaviour in the initial bar growth phase, however, it saturates to a constant value towards the later bar evolutionary phase. 
\par
The most striking finding of this work is that in none of our barred models, the extent of the dark gaps is associated with the bar resonances (corotation, Inner Lindblad resonance, and 4:1ultra-harmonic resonance), as opposed to earlier studies in the literature \citep{Krishnaraoetal2022,Aguerrietal2023}. We checked the bar pattern speed values for our models are well below ($\sim$ by a factor of 2) than those reported in \citet{Krishnaraoetal2022} and \citet{Aguerrietal2023}. We mention that the underlying mass model (through the rotation curve) and the bar pattern speed together set the locations of different resonances associated with the bar. Therefore, the findings presented here, clearly imply that the locations of the dark gaps are not \textit{universally} associated with any of the resonances (of the bar), and depends on both the underlying mass model and the measured bar pattern speed. Furthermore, the bars in all our thin+thick models are slow rotators, that is, $\mathcal{R} (= R_{\rm CR}/ R_{\rm bar}) > 1.4$. We checked that most (about 90 percent) of the MaNGA barred samples used in \citet{Aguerrietal2023} are fast rotators that is, $\mathcal{R} (= R_{\rm CR}/ R_{\rm bar}) < 1.4$ and only about 10 percent qualify as slow rotator ($\mathcal{R} >1.4$). Similarly, in \citet{Krishnaraoetal2022}, most of the barred galaxies (for which the values of $R_{\rm CR}$ and $R_{\rm UHR}$ were reasonably measured) tend to qualify as fast rotators \citep[within the large uncertainties with the corotation radius estimates; see discussions in section 4.1 of][]{Krishnaraoetal2022}. Therefore, the question remains as to whether the association of the dark gap with bar resonances depends on different regimes of bars (i.e. slow versus fast). While our systematic study, as presented here, deals with slow bars, however such a systematic study, dealing with fast bars, is largely missing in the literature, and will be worth pursuing.

\section{Summary and future prospects}
\label{sec:summary}
%&&&&&&&&&&&&&&&&&&&
In summary, we investigated the formation and the subsequent temporal evolution of the bar-induced dark gaps (along the bar minor axis) and their dynamical connection with the bar. We further examined the correlation between the properties of the dark gaps and the bar. We made use of a suite of $N$-body models of thin+thick discs (with varying thick disc mass fraction and different thin-to-thick disc scale length ratios); thereby allowing us to examine the formation and evolutionary trajectory of the dark gaps under diverse dynamical scenarios. Our main findings are listed below.\\

\begin{itemize}

\item{A prominent bar always drives the generation of a dark gap along the bar minor axis. The strength of the dark gap, $\Delta \mu_{\rm max}$ is strongly correlated with the strength of the bar, and this holds for all thin+thick models with varied geometric configuration. Similarly, the length of dark gaps is seen to remain strongly correlated with the bar length, provided a uniform definition is applied in both cases.}

\item{The formation and subsequent growth of dark gaps lead to substantial mass re-distribution along the bar minor axis. For stronger dark gaps (and hence, for stronger bars), the mass loss along the bar minor axis can reach up to $\sim 60-80$ percent of the initial mass contained within the bar extent.}

\item{In all our thin+thick models, the ratio of $R_{\rm CR}$ and bar length, $R_{\rm bar}$ remain above 1.4; thereby qualifying them as slow rotators ($\mathcal{R} > 1.4$). Furthermore, we did not find any robust and \textit{universal} association of the location of dark gaps with the 4:1 ultra-harmonic resonance or the 2:1 Inner Lindblad resonances in any of our thin+thick models, in contrast with earlier studies.}

\end{itemize}

To conclude, our systematic study demonstrates that the properties (strength and extent) of the dark gaps can be used as a robust proxy for the bar properties (strength and extent). 
We mention that the thin+thick models used here do not contain any interstellar gas. The presence of a (dynamically) cold component, such as interstellar gas, makes the disc more susceptible to gravitational instabilities \citep[e.g. see][]{JogandSolomon1984,Jog1996,Bertin2000}. Therefore, it would be worth investigating the secular evolution of bar-induced dark gaps in presence of the interstellar gas.

\section*{Acknowledgement}
%&&&&&&&&&&&&&&&&&&&&&&&&&&&&&&&&&&&&&&&&&&&&&&

We thank the anonymous referee for useful comments which helped to improve this paper. S.G. acknowledges funding from the Alexander von Humboldt Foundation, through Dr. Gregory M. Green's Sofja Kovalevskaja Award. This work has made use of the computational resources obtained through the DARI grant A0120410154 (P.I. : P. Di Matteo). D.A.G. and F.F. were supported by STFC grants ST/T000244/1 and ST/X001075/1. V.C. acknowledges the support provided by ANID through 2022 FONDECYT postdoctoral research grant no. 3220206.

\section*{Data Availability}
\noindent
The simulation data underlying this article will be shared on request to P.D.M (paola.dimatteo@obspm.fr).

%&&&&&&&&&&&&&&&&&&&&&&&&&&&&&&&
\bibliography{my_ref}{}
\bibliographystyle{mnras}
%&&&&&&&&&&&&&&&&&&&&&&&&&&&&&&&

%%%%%%%%%%%%%%%%% APPENDICES %%%%%%%%%%%%%%%%%%%%%

\appendix

\section{Correlation between properties of bars and dark gaps}
\label{appen_correlation_darkgapLengthStrength_allmodels}
%&&&&&&&&&&&&&&&&&&&&&&&

%================================
% Begin figure
%================================
\begin{figure*}
\includegraphics[width=0.85\linewidth]{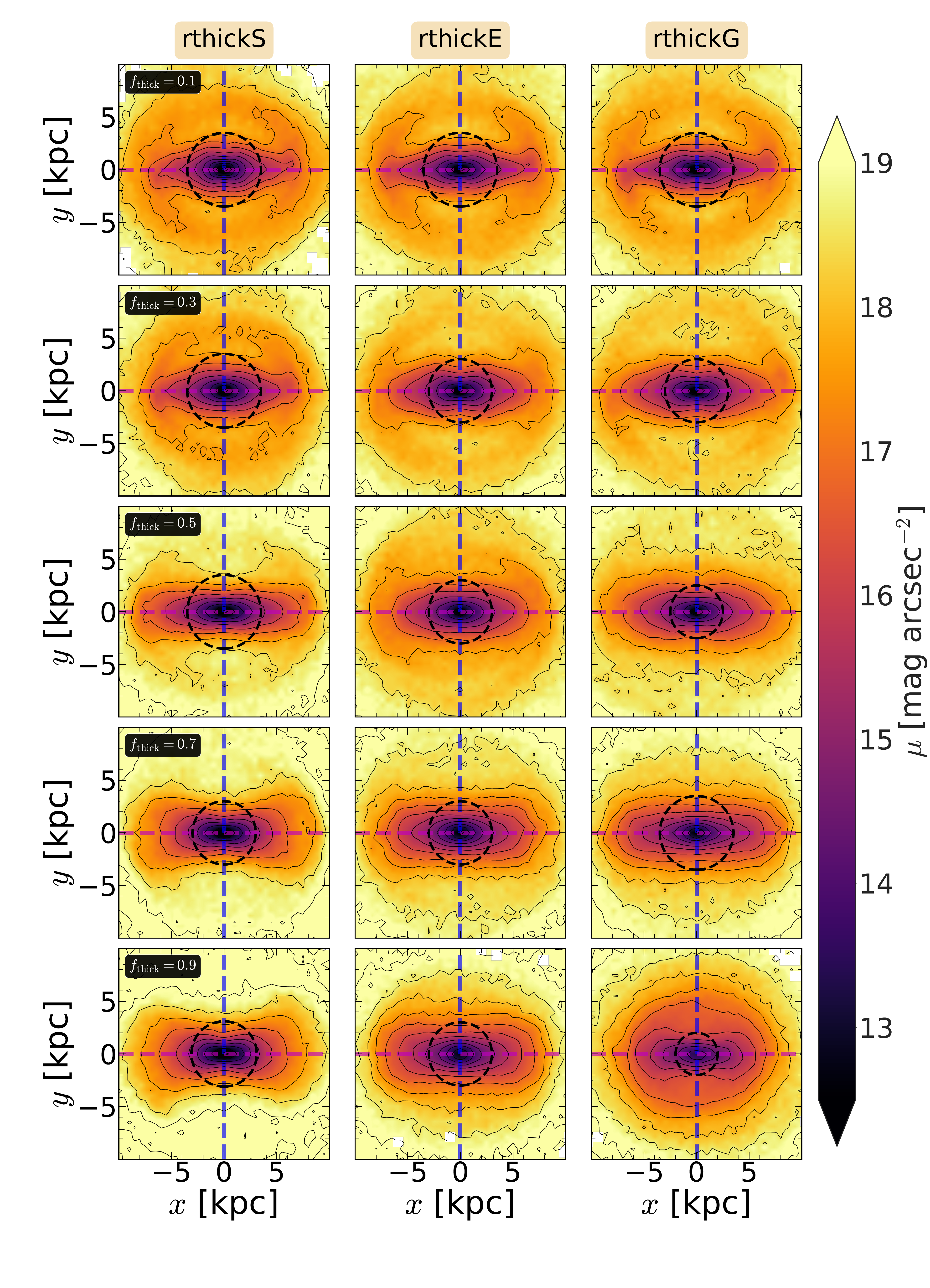}
\caption{Face-on surface brightness distribution, calculated at the end of the simulation run ($t = 9 \Gyr$), for all thin+thick models considered here.
Black solid lines denote the contours of constant surface brightness. For each case, the bar is placed along the $x$-axis. The magenta and the blue dashed lines denote the bar major and minor axis, respectively. The black dashed circle denotes the location of maximum brightness contrast ($\Delta \mu_{\rm max}$), for details see the text. \textit{Left panels} show the surface brightness distribution for the rthickS models whereas  \textit{middle panels} and \textit{right panels} show the surface brightness distribution for the rthickE  and rthickG models, respectively. The thick disc fraction ($f_{\rm thick}$) varies from 0.1 to 0.9 (top to bottom), as indicated in the left-most panel of each row. A magnitude zero-point ($m_0$) of $22.5$ mag arcsec$^{-2}$ and  $\Upsilon_{T}/\Upsilon_{t}$ =1.2 are used to create the surface brightness from the intrinsic particle distribution. Here, 1 arcsec = $1 \kpc$.}
\label{fig:densmap_endsteps_appendix}
\end{figure*}
%================================
% End figure
%================================

%================================
% Begin figure
%================================
\begin{figure*}
\includegraphics[width=\linewidth]{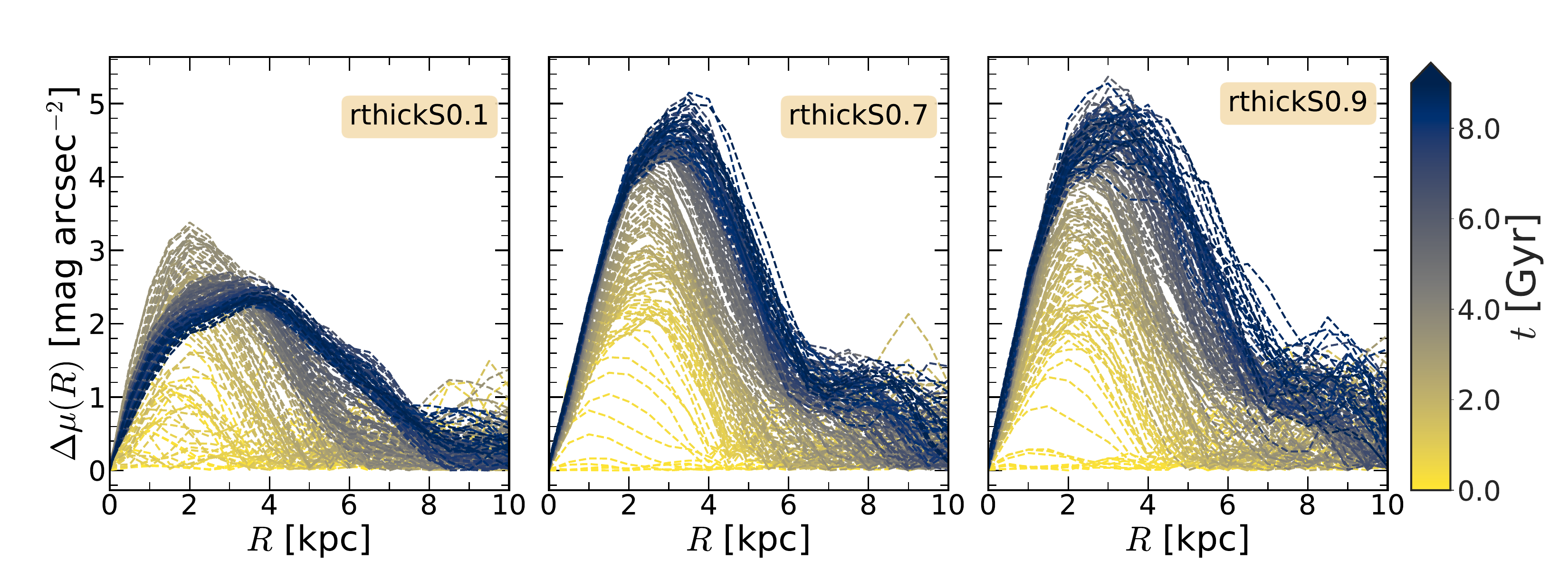}
\caption{Radial variation of $\Delta \mu$, as a function of time (see the colour bar), for three thin+thick models, namely, rthickS0.1, rthickS0.7, and rthickS0.9. A magnitude zero-point ($m_0$) of $22.5$ mag arcsec$^{-2}$ and  $\Upsilon_{T}/\Upsilon_{t}$ =1.2 are used to create the surface brightness from the intrinsic particle distribution. Here, 1 arcsec = $1 \kpc$.}
\label{fig:deltamu_radial_profiles_appendix}
\end{figure*}
%================================
% End figure
%================================

Fig.~\ref{fig:densmap_endsteps_appendix} shows the face-on surface brightness distribution of all 15 thin+thick models, calculated at the end of simulation run ($t = 9 \Gyr$). Even a mere visual inspection of Fig.~\ref{fig:densmap_endsteps_appendix} reveals the presence of conspicuous dark gaps, along the bar minor axis, for almost all thin+thick models considered here.
\par
In Fig.~\ref{fig:deltamu_radial_profiles_appendix}, we show the radial profiles of $\Delta \mu$ as a function of time, for three thin+thick models, namely, rthickS0.1, rthickS0.7, and rthickS0.9. As seen clearly, the temporal evolution of radial profiles of $\Delta \mu$ show variation across the three thin+thick models considered here. While for the model rthickS0.1, the peak location of $\Delta \mu$ progressively shifts towards outer disc region with time, however, for the model rthickS0.9, the peak location of $\Delta \mu$ does not shift appreciably towards outer disc region with time.

\bsp
\label{lastpage}
%&&&&&&&&&&&&&&&&&&&&&&&&&&&&&&&

\end{document}